\documentstyle[sprocl]{article} 

\def\Journal#1#2#3#4{{#1} {\bf #2}, #3 (#4)} 
 
\def\NPB{{\em Nucl. Phys.} B} 
\def\PLB{{\em Phys. Lett.}  B} 
 
\def\PRD{{\em Phys. Rev.} D} 
 
\def\be{\begin{equation}} 
\def\beq{\be}
\def\ee{\end{equation}} 
\def\eeq{\ee}
\def\bea{\begin{eqnarray}} 
\def\eea{\end{eqnarray}} 

\begin{document} 
 
\title{Models Of Dynamical Supersymmetry Breaking} 
 
\author{ L. Randall } 
 
\address{ Center for Theoretical Physics, \\ Laboratory for Nuclear Science
and Department of Physics, \\Massachusetts Institute of Technology,\\
Cambridge, MA 02139, USA}

 
\maketitle\abstracts{ We review  a class of
models of dynamical supersymmetry breaking,
and give a unified description of these models. 
 } 
 
\section{Introduction}
 It is now known that there are many models of dynamical supersymmetry
breaking.
In this talk, I will discuss a class of models of dynamical supersymmetry
breaking, which I have worked on 
with Csaba Cs\'aki, Rob Leigh, Riccardo Rattazzi,
and Witold Skiba \cite{paper1,paper2,adj}  which  illustrate that supersymmetry can be broken
as a result of arbitrary gauge dynamics. The class of models
we discuss is obtained from the anomaly-free DSB model of Affleck,
Dine, and Seiberg \cite{ads} based on $SU(N)$, $N$ odd,
 by removing gauge generators and decomposing the fields
under the resulting $SU(n_1)\times SU(n_2) \times U(1)$ symmetry.
It is intriguing that
although the dynamics of the elements of this
 set of models is different, the analysis 
of supersymmetry breaking with different gauge dynamics is
nonetheless remarkably similar. We will obtain some insight into
this similarity by introducing an adjoint field and analyzing flat
directions and 
the dual phase of the corresponding model.

Throughout this talk, it should be borne in mind that we
will only be discussing the question of dynamical superymmetry
breaking, and not the issue of how the symmetry breaking is communicated
to the superpartners in the visible sector. No matter how supersymmetry
breaking is communicated, the breaking of supersymmetry should be
associated with some mass scale $\Lambda$, where $V=\Lambda^4$ at the
supersymmetry breaking minimum.  Depending on the
mechanism of communicating supersymmetry breaking
to the visible world, the weak scale might be determined
as $\Lambda^2/M_P$ or $\alpha \Lambda$.
 As far as we can tell, if low-energy
supersymmetry is valid, this requires a new mass scale. It would
be more compelling if this mass scale is of dynamical origin,
rather than introduced by hand,
so the hierarchy of mass scales can be associated with the logarithmic
running of a new gauge coupling.  By studying this class of models,
we hope to establish that there can be many models of supersymmetry
breaking, and that the gauge dynamics can correspond to different
phases. The fact that dynamical supersymmetry breaking can happen
in many theories makes it a more compelling possibility. Also,
with more theories which we understand, it is more likely that
we can find a more compelling model for communicating supersymmetry
breaking to the visible world.  Finally, these models
motivate interesting perurbations to the pure gauge theory,
yielding new insights into supersymmetric gauge theories.

\section{Review}

Before discussing our models, it is useful to establish the context,
and what was known before the last few years.  
 Between about '84 and '94, there were only a few known models.
The ``3-2'' model of Affleck, Dine, and Seiberg \cite{ads} remains in some
aspects the most simple and elegant model
in that the field content is familiar and
relatively straightforward, and it is perturbatively analyzable.
The model is based on the gauge group
 $SU(3)\times SU(2) $, and the field content is  $Q(3,2)$, $\bar u(\bar{3})$,
 $\bar d(bar{3})$, $ L(1,2)$.  This model incorporates
a tree-level superpotential which lifts flat directions
and an instanton-generated superpotential which drives
the fields from the origin.  Together, they determine
a minimum which breaks supersymmetry at finite field values. A second model, which is more subtle to analyze, was based on the gauge group
$SU(5)$ with an antisymmetric tensor 10 and a $\bar 5$ \cite{su5}. This model
can be generalized to $SU(N)$ with an antisymmetric tensor ($N$ odd)
which will be the basis for the models which follow. More recently,
additional models with a mechanism of supersymmetry breaking
similar to that in the ```3-2'' model were identified. In these
models, an instanton or gaugino condensation  generated dynamical superpotential
drives fields from the origin, disallowing the classical 
supersymmetry-preserving minimum.  
The existence of such dynamical superpotentials requires
 $\mu_{matter}<\mu_{gauge}$, where $\mu$ is the Dynkin index. Although there are many such models
which one can construct, 
the  $SU(5)$ model with an antisymmetric tensor
 suggests that other dynamical mechanisms
can be operative. This has been demonstrated by many recent models.

In an intriguing model by Intriligator, Seiberg, and Shenkar \cite{iss},
there is an $SU(2)$ gauge group with a spin 3-2 representation
and  a dimension-four gauge invariant superpotential. Anomaly matching
indicates that the model is confining.  The model then looks
like a dynamical O'Raifeartaigh after confinement.

Another interesting model was constructed by Izawa and Yanagida \cite{iy}
and Intriligator and Thomas \cite{it}. The particular model
is based on the gauge group $SU(2)$ with four doublets and six singlets.
There is a tree-level superpotential which lifts most flat directions
(in particular those involving the doublets).  The quantum modified
moduli space drives the fields from the origin however, preventing
a supersymmetric minimum.

These latter models serve to illustrate that the requirement
$\mu_{matter}<\mu_{gauge}$ is not essential. Models in
a confining phase can also break supersymmetry. The
class of models I studied with Cs\'aki, Leigh, and Skiba
involve product gauge groups. Independently, the gauge factors
correspond to a
phase with a quantum modified moduli space, a smooth confining phase,
or a dual phase.
 However, in conjunction with a tree-level superpotential,
one finds an interplay of the dynamics of the two gauge groups
which yields a supersymmetry breaking minimum. The first model
I discuss, the ``4-3-1'' model, seems to
 to exhibit each of the old and new
mechanisms for breaking supersymmetry discussed above, when
analyzed in various limits.  Moreover, the
exact superpotential exhibits interesting features. I will then
discuss other models in which
   Yukawa couplings and confining or dual phase dynamics
dynamics generates a low energy theory incompatible with
supersymmetry.
It is clear that supersymmetry can be broken with any number
of flavors in the presence of an appropriate superpotential.

In the remainder of this talk, we will describe in more
detail the class of theories based on the gauge group 
$SU(n_1) \times SU(n_2) \times U(1)$.

 We begin with a brief elaboration of our review.  Supersymmetry breaking requires
all $D$ and $F$ terms to vanish, so one needs to find a simultaneous
solution to both the $D$ and $F$ flatness equations.  The $D$-flat
directions are easy to characterize.
 They are described by gauge invariant holomorphic functions of the
fields. One can  then check on the ``moduli space''of $D$-flat directions
 whether the $F$ flatness conditions are 
 satisfied. The basic
 idea in dynamical models of supersymmetry
breaking is that the point[s] where supersymmetry
would be preserved classically (generally the origin)
are forbidden by  nonperturbative gauge dynamics.
 How this happens however is very model dependent.
Rather than summarize all phases, will discuss
each mechanism as it arises.
 
In the ``3-2'' model mentioned above, the  superpotential is
 \beq
W=\lambda Q \overline{D} L+{2 \Lambda_3^7 \over {\rm det}\overline{Q}Q},
\eeq 
where the first term is the tree-level superpotential 
and the second term arises from $SU(3)$ instanton dynamics.
One can also incorporate the quantum modified SU(2) constraint \cite{it}
which would be important if $\Lambda_2$ is much greater than $\Lambda_3$.
However, we will analyze the theory in the limit where $g_3>g_2>\lambda$.
There are $14-8-3=3$ flat directions which are characterized by the
holomorphic gauge-invariant combinations
$Q\bar{U}L$, $Q\bar{D}L$, $det\bar{Q}Q$. It is
 readily checked that these $D$-flat directions
are lifted by the superpotential.

The nonperturbatively generated superpotential
drives the fields to large values, whereas the
tree-level terms stabilize the potential to give
a minimum at finite field value.
One can explicitly minimize the potential to show 
that supersymmetry is  broken since the model
can be analyzed with small $\lambda$ in which case
the field values are large at the minimum and the model
is calculable.  One
 could also guess that
supersymmetry is broken from an argument
given by Affleck, Dine, and Seiberg \cite{ads}.
In this model there are no flat directions
and a   spontaneously broken $U(1)_R$
symmetry
where the $R$-charges are $R(Q)=1$, $R(\overline{U})=R(\overline{D})=0$, $R(L)=-3$.  Since there is no scalar partner for the Goldstone boson (as there
are no flat directions), supersymmetry is broken.

In the Intriligator-Seiberg-Shenker model, there is an
 SU(2) gauge theory with Q in spin 3/2 representation.
The low-energy degree of freedom appears to be
 $u=Q^4$. The classical Kahler potential is
$K_{cl}=Q Q^\dagger=(u u^\dagger)^{1/4}$. However, we 
expect the singularity to be smoothed out at the origin if we
have identified the correct field content.
\[ {K=|\Lambda |^{2}
k(uu^\dagger/ |\Lambda |^{8}) \sim
\cases{ uu^\dagger |\Lambda |^{-6}& for $uu^\dagger \ll  \Lambda^8 $ \cr
(uu^\dagger)^{1/4} & for  $uu^\dagger \gg \Lambda^8. $ \cr}}\]
If one introduces the tree-level superpotential,
\[ {W_{tree}=\lambda u.}\]
one obtains the  scalar potential
\[{V_{eff}=(K_{uu^\dagger})^{-1} |W_u|^2=
(K_{uu^\dagger})^{-1}|\lambda|^2.}\]
implying supersymmetry breaking with 
\[
E\sim |\Lambda|^6 |\lambda|^2 \]
Classically, there is no spontaneous symmetry breaking.
However, quantum mechanically the singularity at the
origin in the Kahler potential is smoothed out and supersymmetry
is spontaneously broken.

 In the previous model, the origin, where supersymmetry would
be preserved, is removed quantum mechanically by the modification to $K$.
One can also remove the supersymmetry breaking point quantum
mechanically due to a quantum modified moduli space, as implemented
by a constraint in the  superpotential. The model of Izawa and Yanagida 
\cite{it} and Intriligator and Thomas \cite{it} is based on
 SU(2) with 4 doublets $Q_i$ and six singlets $S^{ij}$
and superpotential
\[ W=\lambda S M+L\left( Pf M-\Lambda_2^4\right) \]
 It is easy to see that the origin of moduli space, required
by $F_{S_{ij}}=0$ is forbidden by the quantum modified
moduli space.
In both these examples, the theory is confining
and it is essential to be able to analyze the theory in the strongly interacting regime.
This is possible due to our understanding of the infrared physics
due to Seiberg \cite{seiberg}.
 
\section{The $4-3-1$ Model}

The next model I will present, the ``4-3-1'' model \cite{paper1}, incorporates
each of the above three mechanisms when analyzed in particular limits.
The breaking of supersymmetry is in all cases a 
conspiracy among strong infrared
dynamics and perturbations. This model and the ones which follow can
 be obtained from a known model
 of supersymmetry based on $SU(n)$ with $n$ odd and field content
breaking 
\beq
A+(n-4)\bar{n}
\eeq
by removing gauge generators. Although this guarantees
an anomaly-free chiral model, it is not guaranteed that
all flat directions are removed \cite{dnns}. It appears nonetheless
to be a good place to look for new models. A better understanding
of this procedure is obtained by adding an adjoint field,
as we will discuss at the end of this talk.

Our first model I discuss is the $4-3-1$ model, based on the gauge
group $SU(4)\times SU(3) \times U(1)$, with field content
 \be 
 A^{\alpha\beta}(6,1)_6, \,  \bar{Q}_a(1,\bar{3})_{-8}, \,
 T^{\alpha a}(4,3)_{-1}, \, \bar{F}_{\alpha I}(\bar{4},1)_{-3}, \,
 \bar{Q}_{a i}(1,\bar{3})_4.
\ee
We include a renormalizable classical superpotential which suffices
to lift all flat directions:
 \begin{eqnarray}
  \label{eq:superpot}
  W_{cl}  &  = &  A^{\alpha\beta} \bar{F}_{\alpha 1} \bar{F}_{\beta 2} +
          T^{\alpha a} \bar{Q}_{a 1} \bar{F}_{\alpha 1} +
          T^{\alpha a} \bar{Q}_{a 2} \bar{F}_{\alpha 2} +
          T^{\alpha a} \bar{Q}_{a 3} \bar{F}_{\alpha 3} + \nonumber \\
      & & \bar{Q}_a \bar{Q}_{b 2} \bar{Q}_{c 1} \epsilon^{abc}. \nonumber
\end{eqnarray}
 This theory also preserves an anomaly-free
 $R$-symmetry, and is therefore a good candidate
  for supersymmetry breaking.
It is educational to investigate the limit
  $\Lambda_3 \gg \Lambda_4$ .
 \begin{eqnarray}
W&=& \bar{b}^3 + \gamma A^{\alpha \beta} \bar{F}_{\alpha 1} \bar{F}_{\beta 2}
   + \lambda_1 F^\alpha_1 \bar{F}_{\alpha 1} + 
     \lambda_2 F^\alpha_2 \bar{F}_{\alpha 2} +
     \lambda_3 F^\alpha_3 \bar{F}_{\alpha 3} + \nonumber \\
 & & \frac{1}{\Lambda_3^5} \left( \bar{F}_{\alpha 4} F^\alpha_i \bar{b}^i
     - {\rm det} F^\alpha_i \right) \nonumber
\end{eqnarray}
where $F_4$ and $\bar{F}_4$ are new composite states.
We see that the 
Yukawa couplings  become mass terms when $SU(3)$ confines.
So the low-energy theory only has one flavor transforming
under the $SU(4)$ group. This means there is
 a dynamical superpotential associated with the SU(4) dynamics.
 \[
  W=\bar{b}^3+\frac{1}{\Lambda_3^5} M_{44} \bar{b}^4 + 
    \left( \frac{\tilde{\Lambda}_4^5}{{\rm Pf}A \, M_{44}}\right)^{\frac{1}{2}},
\]
where $M_{44}=\bar{F}_{\alpha 4} F_4^\alpha$.
This limit is interesting because we can identify two sources of 
supersymmetry breaking. The term
 $\bar{b^3}$ permits an ISS-like mechanism due to the confined Yukawa term. There is also a 
 competition between the SU(3) and SU(4) generated terms which
acts like the mechanism of the $3-2$ model, although
the role of the tree level superpotential is  performed by
a dynamical term. One can also investigate supersymmetry
breaking in the opposite limit, in which $\Lambda_4$ is strong
first. In that case, the quantum
  quantum modified constraint seems to be the source
of supersymmetry breaking. However,
  in either case some terms are neglected. It is therefore
 safest to look at the exact vacuum, taking
 both $\Lambda_3$ and $\Lambda_4$ strong from the beginning
so that we can investigate the true low-energy state.
 
 Using  holomorphy, various limits,
and the constraints associated with these limits,
one can derive the  exact superpotential. Very strong
among these   the requirement of reproducing correctly 
  the constraints in the
limits $\Lambda_3 \gg \Lambda_4$ and $\Lambda_4 \gg \Lambda_3$.
 It is interesting to see how
the classical constraints are reproduced (sometimes
with quantum modifications) by equations
of motion of this superpotential, which often occurs
in nontrivial ways.

When deriving the superpotential, we found it most
useful
  to work in terms of
$U(1)$ charged holomorphic invariants  which describe
the low energy spectrum if SU(3) and SU(4) are confining:
\begin{eqnarray}
\label{eq:classical}
  M_{iI}= & T^{\alpha a} \bar{Q}_{a i} \bar{F}_{\alpha I}        & 0   
  \nonumber \\
  M_{4I}= & T^{\alpha a} \bar{Q}_{a} \bar{F}_{\alpha I}          & -12 
  \nonumber \\
  X_{IJ}= & A^{\alpha\beta} \bar{F}_{\alpha I} \bar{F}_{\beta J} & 0  
  \nonumber \\
  X_{I4}= & \frac{1}{6} A^{\beta \alpha} \bar{F}_{\beta I} 
            \epsilon_{\alpha \gamma \delta \zeta} T^{\gamma a}
            T^{\delta b} T^{\zeta c} \epsilon_{a b c}            & 0   
  \nonumber \\
 {\rm Pf}A=& \epsilon_{\alpha \beta \gamma \delta} 
            A^{\alpha \beta} A^{\gamma \delta}                   & 12 
\nonumber  \\
  Y_{ij}= & \epsilon_{\alpha \beta \gamma \delta} 
            A^{\alpha \beta} T^{\gamma a} \bar{Q}_{ai}
            T^{\delta b} \bar{Q}_{b j}                           & 12  
   \nonumber \\
  Y_{i4}= & \epsilon_{\alpha \beta \gamma \delta} 
            A^{\alpha \beta} T^{\gamma a} \bar{Q}_{ai}
            T^{\delta b} \bar{Q}_{b}                             & 0   
   \nonumber \\
  \bar{B}=& \frac{1}{6} \bar{F}_{\alpha I} \bar{F}_{\beta J} 
            \bar{F}_{\gamma K} \epsilon^{I J K}
            T^{\alpha a} T^{\beta b} 
            T^{\gamma c} \epsilon_{a b c}                        & -12 
  \nonumber \\
  \bar{b}^i=& -\frac{1}{2} \bar{Q}_a \bar{Q}_{b j} \bar{Q}_{c k}
            \epsilon^{i j k} \epsilon^{a b c}                    & 0   
   \nonumber \\
  \bar{b}^4=& \frac{1}{6} \bar{Q}_{a i} \bar{Q}_{b j}
            \bar{Q}_{c k} \epsilon^{i j k} \epsilon^{a b c}      & 12  
   \nonumber 
\end{eqnarray}
We have given the bound states and their $U(1)$ charges.

 There are a number of classical constraints, which are
\begin{eqnarray}
\label{chargedconst}
  & & 4 \, X_{I4}X_{JK}\epsilon^{IJK}-\bar{B} \, {\rm Pf}A=0 \nonumber \\
   & & \epsilon^{ijk} \epsilon^{IJK} 
     \left( {\rm Pf}A \, M_{iI} M_{jJ} M_{kK} - 6 Y_{ij} M_{kI} X_{JK} \right) 
   = 0 \nonumber \\
 & & \epsilon^{ijk} \epsilon^{IJK} 
     \left({\rm Pf}A \, M_{4I} M_{jJ} M_{kK} - 2 Y_{jk} M_{4I} X_{JK}
     + 4 Y_{j4} M_{kI} X_{JK} \right) = 0 \nonumber \\ 
  & & Y_{i4}\bar{b}^i=0 \nonumber \\
  & & \bar{B}\bar{b}^4-\frac{1}{6} \epsilon^{ijk}\epsilon^{IJK}M_{iI}M_{jJ}
      M_{kK}=0 \nonumber \\
  & & \bar{B}\epsilon^{kij}Y_{ij}-2 \, \epsilon^{kij}\epsilon^{IJK}
      M_{iI}M_{jJ}X_{K4}=0 \nonumber \\
  & & M_{4I}\bar{b}^4+M_{iI}\bar{b}^i=0 \nonumber \\
  & & \epsilon^{ijk}Y_{jk}M_{4I}+2 \, \epsilon^{ijk}M_{jI}Y_{k4}+
      4 \, X_{I4}\bar{b}^i=0 \nonumber \\
  & & \epsilon^{IJK}\epsilon^{ijk}M_{iI}M_{jJ}M_{4K}Y_{k4}=0.
\end{eqnarray}
 
To derive the quantum theory, it is useful 
to observe that 
from the vantage point of SU(4), the theory contains
an antisymmetric tensor and three flavors, and therefore
has a quantum modified contraint \cite{ptanti}   $4X_{I4}X_{JK} \epsilon^{IJK}-\bar{B}PfA=\Lambda_4^8$,
whereas from the point of view of SU(3), there
are four flavors and the theory is confined with no modification
to the classical constraints.
It is therefore easiest to derive the superpotential by
first taking the  $\Lambda_3\gg \Lambda_4$ limit,
since this permits the full $SU(4)$ symmetry to remain unbroken
in the $SU(3)$ moduli space. One can then
incorporate the constraints from $SU(4)$ with {\it four} flavors
(one additional flavor arising in the confined SU(3) theory).
\begin{eqnarray}
\label{exsuppot}
W & \hspace{-4pt} = & \hspace{-4pt} \bar{b}^3 +X_{12} + M_{11} + M_{22}
                    + M_{33} + \frac{1}{\Lambda_3^5} \left( M_{i4} \bar{b}^i 
                    - B \right) + \nonumber \\
 \hspace{-4pt}& & \hspace{-4pt}
                  \frac{1}{24 \, \Lambda_3^5\Lambda_4^8} 
                  \Big( 24 \, BX_{IJ}X_{KL}\epsilon^{IJKL}+ 
                  6 \, \bar{B}Y_{ij}Y_{kl}\epsilon^{ijkl}
                  -24 \, B\bar{B}{\rm Pf}A + \nonumber \\
 \hspace{-4pt}& & \hspace{-4pt} 
     {\rm Pf}A \epsilon^{ijkl} \epsilon^{IJKL} M_{iI} M_{jJ} M_{kK} M_{lL}
     - 12 \, \epsilon^{ijkl} Y_{ij} M_{kI} M_{lJ}
X_{KL}\epsilon^{IJKL} \Big) \nonumber
\end{eqnarray}
 The form of this superpotential is 
 determined by symmetries and $\Lambda_3\gg \Lambda_4$ and  the $\Lambda_4 \gg \Lambda_3$ limits. The relative coefficient
is determined by the second limit, as can be seen by addressing
two questions.

There are however a couple of things
which seem strange at first.
The first apparent puzzle was what happened to the $SU(4)$ quantum
modied constraint. The second puzzle is what happens
to the many constraints of the original classical theory?
It  turns out  these puzzles are resolved in an interesting
manner through the equations of motion associated with
classically vanishing fields.
 
First consider the equation of motion associated with 
 $B=T\bar{Q}T \bar{Q} T\bar{Q} T\bar{Q}$:
 which is
\[
4X_{I4}X_{JK} \epsilon^{IJK}-\bar{B}PfA=\Lambda_4^8
\]
This  gives us  precisely  the SU(4) quantum modified constraint,
which is how a relative coefficient in the above superpotential
was determined.

Next consider the equation of motion associated
with the field, $M_{i4}=T \bar{Q}_i T T T $:
   \[ \epsilon^{ijk} \epsilon^{IJK} 
     \left( {\rm Pf}A M_{iI} M_{jJ} M_{kK} - 6 Y_{ij} M_{kI} X_{JK} \right) 
   = 6 \Lambda_4^8 \, \bar{b}^4 \]
  \[ \epsilon^{ijk} \epsilon^{IJK} 
     \left({\rm Pf}A M_{4I} M_{jJ} M_{kK} - 2 Y_{jk} M_{4I} X_{JK}
     + 4 Y_{j4} M_{kI} X_{JK} \right) = 2 \Lambda_4^8 \, \bar{b}^i 
      \]
This equation means that we can solve for the fields $b_i$.
Many of the constraint equations are automatically satisfied
by this solution.  This equation can also
 be interpreted as a quantum modified constraint, this
  when one of the flavors of the confined SU(3)
theory (four flavors) is given a mass through   a nonvanishing $b_i$.
By matching the scale $\Lambda_3$ through the mass proportional
to $b_i$, one derives precisely this constraint.
 
The full list of quantum modified constraints is 
\begin{eqnarray}
\label{qc1}
  & & \hspace{-1cm} 4 \, X_{I4}X_{JK}\epsilon^{IJK}-\bar{B}{\rm Pf}A=
      \Lambda_4^8 \nonumber  \\
   \label{qc2}
  & & \hspace{-1cm} \epsilon^{ijk} \epsilon^{IJK} 
     \left( {\rm Pf}A M_{iI} M_{jJ} M_{kK} - 6 Y_{ij} M_{kI} X_{JK} \right) 
   = 6 \Lambda_4^8 \, \bar{b}^4 \nonumber \\
\label{qc3}
  & & \hspace{-1cm} \epsilon \epsilon
     \left({\rm Pf}A M_{4I} M_{jJ} M_{kK} - 2 Y_{jk} M_{4I} X_{JK}
     + 4 Y_{j4} M_{kI} X_{JK} \right) = 2 \Lambda_4^8 \, \bar{b}^i \nonumber \\
\label{qc4}
  & & \hspace{-1cm} \epsilon^{IJK}\epsilon^{ijk}M_{iI}M_{jJ}M_{4K}Y_{k4}=
      2 \, B M_{4I} X_{JK} \epsilon^{IJK} \nonumber \\
\label{qc5}
  & & \hspace{-1cm} \bar{B}\epsilon^{kij}Y_{ij}-2 \, \epsilon^{kij}
      \epsilon^{IJK}
      M_{iI}M_{jJ}X_{K4}=-2\,M_{i4}M_{jI}\epsilon^{kij}X_{JK} \epsilon^{IJK}\nonumber
\end{eqnarray}

 Each classical constraint has associated equation of motion
in quantum theory
 and vice versa. Observe that the quantum modifications
can be
  field dependent. The classcial limit in each case is reproduced
either because a field 
 or $\Lambda$ goes
to zero in  the classical limit.

  Given the exact theory, it can  be shown that the above superpotential does not permit
a supersymmetric minimum, and is in this sense
 of O'Raifeartaigh type. The proof is complicated but doable.
  The above model can be generalized to other models
based on the gauge group  $SU(n)\times SU(3)\times U(1)$.
Here $SU(n)$ confines, resulting in heavy flavors of $SU(3)$.
When these are integrated out, one can show the remaining
dynamical superpotential forbids supersymmetry preservation.

 It might be thought that a ``dynamical'' superpotential
proportional to inverse powers of the field
or  a quantum modified superpotential
required to forbid the origin where the classical supersymmetric
solution
occurs. However, the
   $SU(n)\times SU(3)\times U(1)$ generalizations
of the $4-3-1$ model suggest this
is false.  In fact, one 
 can have models based on confining potential with {\it
no}
quantum modification. 
  Moreover, models can have   both groups
in the conformal window.  These also break supersymmetry!
 The mechanism  is analagous to that described above.
 \newcommand{\drawsquare}[2]{\hbox{%
\rule{#2pt}{#1pt}\hskip-#2pt
\rule{#1pt}{#2pt}\hskip-#1pt
\rule[#1pt]{#1pt}{#2pt}}\rule[#1pt]{#2pt}{#2pt}\hskip-#2pt
\rule{#2pt}{#1pt}}
\newcommand{\Yfund}{\raisebox{-.5pt}{\drawsquare{6.5}{0.4}}}
\newcommand{\Ysymm}{\raisebox{-.5pt}{\drawsquare{6.5}{0.4}}\hskip-0.4pt%
        \raisebox{-.5pt}{\drawsquare{6.5}{0.4}}}
\newcommand{\Yasymm}{\raisebox{-3.5pt}{\drawsquare{6.5}{0.4}}\hskip-6.9pt%
        \raisebox{3pt}{\drawsquare{6.5}{0.4}}}

\section{Confining Models}

 We first present a confining model based on the
gauge group  $SU(n)\times SU(4)\times U(1)$ \cite{paper2}, again obtained
by decomposing an $SU(n+4)$ ADS model with an antisymmetric tensor.
 We can demonstrate supersymmetry breaking in a model where
 $SU(n)$ confines with no quantum modification to constraints
and  $SU(4)$  is in the  conformal window. The field content is
 \begin{eqnarray}
 &  & A(\Yasymm,1)_8 + 
                       a(1,\Yasymm)_{-2 n} +
                       T(\Yfund,\Yfund )_{4-n} \nonumber \\
 &  & \bar{F}_{I} 
       (\overline{\Yfund},1)_{-4} + \bar{Q}_{i} (1,\overline{\Yfund})_n \nonumber
\end{eqnarray}
and the superpotential is
\begin{eqnarray}
\label{wtree}
W_{tree}&=&A\bar{F}_1\bar{F}_2+\ldots +A\bar{F}_{n-2}\bar{F}_{n-1}\\
&&+a\bar{Q}_2\bar{Q}_3+\ldots +a\bar{Q}_{n-1}\bar{Q}_1 +\nonumber \\
& & T\bar{F}_1\bar{Q}_1+\ldots +T\bar{F}_n\bar{Q}_n.
\end{eqnarray}
 It  can be shown  that flat  directions are lifted. Note the shift in indices;
a better undrstanding of this superpotential was given in Ref. \cite{adj}
where it was shown that this is just one of the generic set of superpotentials
which suffice to preserve flat directions. This superpotential
is special in that it preserves an $R$-symmetry. However,
in Ref. \cite{adj}, it was shown that this is in fact not necessary.
  However, if $f_{ij} A \bar{F}_i \bar{F}_j+q_{ij}a \bar{Q}_i \bar{Q}_j$,
flat directions will not be lifted if 
$f$ is proportional  to $q$  which is why simply taking the
same Yukawa couplings would not work.

This theory can be analyzed
   at large  $\Lambda_n$.
 For these models,  $SU(n)$ is confining
 with SU(n) invariants
\begin{eqnarray}
\label{suninvariants}
&& X_{IJ}=A^{\alpha \beta}\bar{F}_{\alpha I}\bar{F}_{\beta J} \nonumber \\
&& \bar{B}=\bar{F}_{\alpha_1 1}\ldots \bar{F}_{\alpha_n n}
\epsilon^{\alpha_1 \ldots \alpha_n} \nonumber \\
&& (B_1)^a=T^{\alpha_1 a} A^{\alpha_2 \alpha_3}\ldots 
A^{\alpha_{n-1} \alpha_n} \epsilon_{\alpha_1 \ldots \alpha_n} \nonumber \\
&& (B_3)_a=\epsilon_{abcd} T^{\alpha_1 b}T^{\alpha_2 c}T^{\alpha_3 d}
 A^{\alpha_4 \alpha_5}\ldots 
A^{\alpha_{n-1} \alpha_n} \epsilon_{\alpha_1 \ldots \alpha_n} \nonumber \\
&& M_I^a=T^{\alpha a}\bar{F}_{\alpha I}\nonumber
\end{eqnarray}
plus the $SU(n)$ singlets $a$ and $\bar{Q}_i$.
 Now we consider the resulting $SU(4)$ theory.
There are   $n+1$ flavors of SU(4), with one additional flavor
arising in the confined spectrum. In the presence of the Yukawa couplings,
one
 can integrate out $n$ flavors since $M$ and $\bar{Q}$ are massive.
The remaining theory has only one flavor of $SU(4)$ and
there is a dynamical superpotential.
 \begin{eqnarray}
\label{sun}
W&=&
X_{12}+\ldots +X_{n-2,n-1}+\frac{1}{\Lambda_n^{2n-1}}\bar{B}m+
\left[\frac{\tilde{\Lambda}_4^5}{{\rm Pf}a \,
m}\right]^{\frac{1}{2}}\nonumber
\end{eqnarray}
where ${\rm Pf}a=a^{ab}a^{cd}\epsilon_{abcd}$, $m=B_1^aB_{3 a}$, and
$\tilde{\Lambda}_4$ is the dynamical scale of the effective one flavor
$SU(4)$ theory and we have integrated out heavy flavors.
 It is clear that there is no supersymmetric solution to the equations of motion.

\section{Dual Phase Analysis}
 
We now describe a model with a dual description of 
the dynamics, which will generate a remarkably similar
low-energy theory to the previous confining model.
The class of models is again derived by removing generators
from the ADS models; here we investigate the theories with gauge group
  $SU(n) \times SU(5)
\times U(1)$ ($n$ even) \cite{paper2}. The field content is 
 \begin{eqnarray}
\Yasymm & \rightarrow & A(\Yasymm,1)_{10} + 
                       a(1,\Yasymm)_{-2n} +
                       T(\Yfund,\Yfund )_{5-n} \nonumber \\
(n+1)\cdot \, \overline{\Yfund} & \rightarrow & \bar{F}_{I} 
       (\overline{\Yfund},1)_{-5} + \bar{Q}_{i} (1,\overline{\Yfund})_n \nonumber
\end{eqnarray}
where $i,I=1,\ldots,n+1$. 

The tree-level superpotential is
\begin{eqnarray}
\label{tree651}
W_{tree}&=&A\bar{F}_1\bar{F}_2+
\ldots+A\bar{F}_{n-1}\bar{F}_{n}
+a\bar{Q}_2\bar{Q}_3+
\ldots+a\bar{Q}_{n}\bar{Q}_1 +\nonumber \\
& & T\bar{F}_1\bar{Q}_1+\ldots +T\bar{F}_{n+1}\bar{Q}_{n+1} \nonumber
\end{eqnarray}
and lifts all flat directions.

The $SU(5)$ gauge group has an antisymmetric tensor and $n$
flavors while the $SU(n)$ has an antisymmetric tensor and five
flavors. 
 The $SU(5)$ and $SU(n)$ groups are  in the  conformal phase. The theory
is therefore best analyzed in a dual phase.
 Since we  better understand the dual for odd $n$, we dualize SU(5) \cite{pouliot}.
    As before, some of the tree-level Yukawa terms are
mapped into mass terms in the dual description. 
  After integrating out these
massive fields, we have the   effective superpotential:
\[
W=H_{23}+H_{45}+H_{n1}+ H l^2 + \bar{a} x^2 + \tilde{M} p +
    \frac{\tilde{\Lambda}_n^{n+1}}{(\tilde{M} \tilde{X}^{(n-4)/2} {\rm Pf}\,A)^{1/2}}  
\]
where $\tilde{M}=B_1 \bar{q}$, $\tilde{X}= A\bar{q}\bar{q}$ and
${\rm Pf}\,A=A^{n/2}$, while $\tilde{\Lambda}_n$ is the effective $SU(n)$ scale. 
    One can observe that it looks  very much like the confining case;
the additional gauge groups appear to be simply spectators.
   In both the  confining and dual cases, the dynamics of one
gauge group introduces extra flavors of the other gauge group.
  In both cases, some tree level terms are mapped into mass terms
due to dynamical effects of this group
  After integrating out the  massive flavors,   only a single
flavor remains besides the antisymmetric tensor.
 In the dual example vs. the confining one, we
see   $\tilde{M}$ plays the
role of $m$ and p the role of $\bar{B}$. 
  A dynamically generated superpotential is produced
which together with the remainder of the superpotenial generated
by the first group breaks supersymmetry.

\def\Title#1#2{\rightline{#1}
\ifx\answ\bigans\nopagenumbers\pageno0\vskip1in%
\baselineskip 15pt plus 1pt minus 1pt
\else
\def\listrefs{\footatend\vskip 1in\immediate\closeout\rfile\writestoppt
\baselineskip=14pt\centerline{{\bf References}}\bigskip{\frenchspacing%
\parindent=20pt\escapechar=` \input
refs.tmp\vfill\eject}\nonfrenchspacing}
\pageno1\vskip.8in\fi \centerline{\titlefont #2}\vskip .5in}
%
%
\def\adj{\Sigma}
\def\bF{\bar{F}}
\def\bP{\bar{P}}
\def\brf{\bar{f}}
\def\bp{\bar{p}}
\def\l{\lambda}
\def\tl{\tilde{\l}}
\def\tr{{\rm tr}}
 \def\begi{\begin{itemize}}
\def\endi{\end{itemize}}
\def\begsm{\begin{small}}
\def\endsm{\end{small}}
\def\beq{\[}
\def\eeq{\]}
\def\mpl{M_p}
\def\Mpc{{\rm Mpc}}
\def\la{<}
%
\font\zfont = cmss10 
\font\zfonteight = cmss8 
\def\ZZ{\hbox{\zfont Z\kern-.4emZ}}
\def\sZZ{\hbox{\zfonteight Z\kern-.4emZ}}
\def\RR{\hbox{\zfont  l\kern-.1emR}}
\def\RP{\RR\hbox{\zfont P}} 

%
\def\boxit#1{\vbox{\hrule\hbox{\vrule\kern3pt
\vbox{\kern3pt#1\kern3pt}\kern3pt\vrule}\hrule}}
\newdimen\str
\def\xstrut#1{\dimen\str = #1
\hbox{\vrule height .8dm\str depth .2dm\str width 0pt}}
\def\fboxit#1#2{\vbox{\hrule height #1 \hbox{\vrule width #1
\kern3pt \vbox{\kern3pt#2\kern3pt}\kern3pt \vrule width #1 }
\hrule height #1 }}
\def\yboxit#1#2{\vbox{\hrule height #1 \hbox{\vrule width #1
\vbox{#2}\vrule width #1 }\hrule height #1 }}
\def\fillbox#1{\hbox to #1{\vbox to #1{\vfil}\hfil}}
\def\dotbox#1{\hbox to #1{\vbox to 8pt{\vfil}\hfil $\cdots$ \hfil}}
\def\ybox{\yboxit{0.4pt}{\fillbox{8pt}}\hskip-0.4pt}
\def\nbox{\yboxit{0pt}{\fillbox{8pt}}}
\def\yboxs{\yboxit{0.4pt}{\fillbox{5pt}}\hskip-0.4pt}
\def\nboxs{\yboxit{0pt}{\fillbox{5pt}}}
\def\cdotbox{\yboxit{0.4pt}{\dotbox{24pt}}\hskip-0.4pt}
\def\tableaux#1{\vcenter{\offinterlineskip \halign{&\tabskip 0pt##\cr
#1}}\ }
\def\cropen#1{\crcr\noalign{\vskip #1}}
\def\cry{\cropen{-0.4pt}}
\def\asym{\tableaux{\yboxs\cry\yboxs\cry}}
\def\fund{\tableaux{\yboxs\cry}}
\def\afund{\bar{\fund}}
%

 %
\def\npb#1#2#3{Nucl. Phys. {\bf B#1} (#2) #3}
\def\plb#1#2#3{Phys. Lett. {\bf #1B} (#2) #3}
\def\prd#1#2#3{Phys. Rev. {\bf D#1} (#2) #3}
\def\prl#1#2#3{Phys. Rev. Lett. {\bf #1} (#2) #3}
\def\mpl#1#2#3{Mod. Phys. Lett. {\bf A#1} (#2) #3}
\def\nl{\noindent}

\section{ADS Model With an Adjoint}
 
In light of these similarities, and all the models
discussed so far broke supersymmetry, it is 
natural to ask whether there exists a unified picture
of these models. With Rob Leigh and Riccardo Rattazzi,
I addressed this question.
  All these models were obtained from the gauge
theory SU(N) ($N$ odd) with $A  + (N-4) \bar{F}$ by removing 
gauge generators; one can obtain these vacua 
  directly in a theory
with a heavy adjoint. With suitable choice of superpotential,
one can generate vacua with gauge symmetry
  $SU(n_1)\times
SU(n_2)\times\ldots\times SU(n_k)\times U(1)^{k-1}$. This
allows us to probe the theories considered previously
as well as a large class of new theories \cite{adj}.
  
 Begin with $SU(N)$ gauge theory with matter
 	\begin{eqnarray}
		\adj&=& {\bf Adj}\\ \nonumber
		A  & =&\; \asym\\ \nonumber
		\bF_I &=&\; \afund,\;\;\; I=1,\ldots,(N-4)\\ \nonumber
	 \end{eqnarray}
   and  a superpotential
\begin{eqnarray}
 	W&=& {a_o\over k+1}\;\tr\adj^{k+1}+ W_k(\adj) + \nonumber  \\
	&& \sum_{m,n=1}^{k}\l_{mn}^{IJ} \bF_I \adj^{m-1}A\adj^{n-1}\bF_J 
 + \ldots \nonumber
\end{eqnarray}
where $W_k$ is a polynomial of degree $k$ in $\adj$.

 For generic $W_k$, the model has several classical vacua on the
Coulomb branch; these are determined by the solutions of
$$a_o \adj^k +W'_k\sim 0,$$ 
The general solution takes the form
\[
\Sigma = \left(\begin{array}{ccccccccc}
a_{1}      \\
   &\ddots\\
&&a_{1}\\[.75ex]
&&& a_2\\
&&&&\ddots\\
&&&&&a_2\\[.75ex]
&&&&&&a_k\\
&&&&&&&\ddots\\
&&&&&&&&a_k
\end{array}\right)
\]
  We see that  $k=2$ gives the models
which were previously discussed, 
where as   $k>2$  gives  $SU(n_1)\times \ldots SU(n_k)\times U(1)^{k-1}$.
  
 To analyze the theory, we found it useful to
first ``deconfine" with    SP((N-3)/2).
  This permits us to think of the low-energy
theory (below the SP confining scale) as a theory
of an adjoint and  $N_F=N-3$ flavors. This is a theory
that we
  know how to dualize. In particular,
we only know the dual with the superpotential!
 This is precisely the superpotential which is required to generate
the desired vacua.

In the end, after integrating out massive flavors, we have 
\[
W=\ldots+g \tilde{\mu}\bar{p} p+{\Lambda_m^{m+2} \over \tilde{\mu}^{(m-1)/2}} \left( \tilde{A}^{m-2} p\right)^{1/2}
\]
 Notice the similarity to the previous individual models!
  But here, the dynamical superpotential is generated by the SP dynamics.
  It was essential that we had only a small number (m-1) flavors of SP(m).
 
 Now let us consider the case $k > 2$.
  We can apply the same procedure to
study $SU(n_1)\times SU(n_2)\times SU(n_3)\times U(1)^2$ (here $k=3$)
 It turns out theory with massive modes integrated
out is in the conformal phase. The
 Lagrangian indicates nothing singular;
there is no apparent reason for supersymmetry breaking.   

Because the theory is in the conformal phase, it is
difficult to deduce a definite conclusion.
However there is strong evidence that these
theories do not break supersymmetry, which 
can be best understood from explicit examples.  The
root of the problem appears to be the 2m-baryonic
operators
 \[
B_{m_i,p_{ij}}=\prod A_i^{m_i} T_{ij}^{p_{ij}}
\]
which exist only for $k>2$.
 
It is clear that these operators
need to be lifted; otherwise they would acquire expectation
value and break the gauge symmetry.   However
it appears to be impossible to lift all such operators
(and the other lower dimensional operators
which one needs to lift) without violating an $R$-symmetry.
 The models have the property that without inclusion of operators
to lift these flat directions in the superpotential,
supersymmetry is broken for finite field value, but there
is a runaway direction.
  With the inclusion of these operators, there is a supersymmetric minimum.
It appears   that  the $k>2$ models are really different.
This also demonstrates
 that the  procedure of simply removing generators
to make new models will not  be successful in general,
and that we were lucky to find a class of models which worked. We 
found that adding the adjoint does give some insight into
the models which interested us; however even that
is not always conclusive and more explicit methods appeared 
to be required.  Other interesting conclusions had to do 
with the role of $R$-symmetry in these models. There are
many cases where an anomaly-free $R$-symmetry
was not necessary, but only an effective $R$-symmetry.
However, it is difficult to make overly broad generalizations
which precisely specify when such a symmetry is required.
Some insights into these points is presented in Ref. \cite{adj}.

The last couple of years has witnessed a dramatic increase
in our understanding of strong dynamics, and of supersymmetry
breaking.
It is now clear
that spontaneous
breaking of supersymmetry
 does not require $\mu_{matter}<\mu_{gauge}$.

In this talk, we explored  a particular class of models.
We found that in the $4-3-1$ model, there
were various mechanisms at work, and that
we could also understand the exact theory.
In the case where one factor
of the product group was confining (without
quantum modification),
we found that there was
nonetheless supersymmetry breaking
which could be understood
after integrating out the heavy confined fields.
 In the  conformal phase,
we observed the same phenomena, except the masses
were attributable to matching the Yukawa terms
onto terms involving the dual meson fields.
That these models  break supersymmetry can be understood by
addition of a massive adjoint field.
Notice that index arguments did not suffice; new vacua
are present in this theory.
 However using deconfinement (antisymmetric tensor) and 
Kutasov duality we could analyze the entire class of models
at once. Again, we found a similar mechanism
but the  dynamical superpotential is a 
result of the additional gauge group which confines
fundamentals  into an antisymmetric tensor. It should
be noted that the proof required a rigorous analysis
of flat directions to be valid.

On the other hand, we found there was no obvious
mechanism for supersymmetry breaking 
when there were three or more nonabelian factors (or
an additional U(1)). In fact, the dual we constructed
following an analogous procedure was more strongly coupled.
That these theories do not break supersymmetry
 seems to be due to new flat directions present in this theory.

One major lesson from this class of models remains
that even theories with large numbers of
flavors can break supersymmetry. It is not necessary that
there is
an  instanton generated superpotential, 
gaugino condensation, or a quantum modification in initial
formulation of the theory.
 Infrared physics sees only the states left massless after confinement
and can nonetheless break supersymmetry.
 Clearly there are many new possibilities.
 We have given only a few to demonstrate that supersymmetry
might be broken even if you are in  different phases.
 By exploring the many  new candidates we might
ultimately find  more compelling mechanisms, both to break
and communicate supersymmetry breaking.

\section*{Acknowledgments} 
  This work is supported in part by DOE under cooperative agreement
\#DE-FC02-94ER40818, NSF Young Investigator Award,
Alfred P. Sloan Foundation Fellowship, DOE Outstanding Junior
Investigator Award.  I would like to acknowlege my collaborators
on the papers on which this talk is based: Csaba Cs\'aki,
Rob Leigh, Riccardo Rattazzi, and Witold Skiba.

\end{document}